\documentclass[conference]{IEEEtran}
\IEEEoverridecommandlockouts
\ifCLASSINFOpdf
\usepackage{graphicx}
\usepackage{epstopdf}
\usepackage{float}
\else
\fi
\usepackage{amsfonts}
\usepackage{amssymb}
\usepackage{amsmath}
\usepackage{multirow}
\usepackage{array}
\usepackage{booktabs}
\usepackage{multirow}
\usepackage{epstopdf,epsfig}

\begin{document}

\title{Tomlinson-Harashima Precoded Rate-Splitting for Multiuser MIMO Systems \vspace{-0.05em}}

\author{Andre R. Flores $^1$, Bruno Clerckx $^2$
and Rodrigo C. de Lamare $^1$,$^3$ \\
$^1$ ~Centre for Telecommunications Studies, Pontifical Catholic
University of Rio de Janeiro, Brazil \\
$^2$ ~Imperial College London, United Kingdom \\
$^3$ ~Department of Electronic Engineering, University of York,
United Kingdom \\
Emails: andre.flores@cetuc.puc-rio.br, b.clerckx@imperial.ac.uk,
delamare@cetuc.puc-rio.br
\thanks{This work is partly funded by the CNPq, CGI and FAPERJ.}\vspace{-0.75em}}

\maketitle

\begin{abstract}
In this work, we investigate the performance of Rate-Splitting (RS)
based on Tomlinson-Harashima Precoding (THP) in a multiple-antenna
broadcast channel with perfect and imperfect Channel State
Information at the Transmitter (CSIT). In particular, we consider RS
using centralized and decentralized THP structures, where only one
user splits its message into a common and private part, and develop
expressions to describe the signal-to-interference-plus-noise (SINR)
ratio and the sum rates associated with these schemes. Furthermore,
we also assess the performance achieved by RS combined with
Dirty-Paper Coding (DPC). Simulations show that RS with THP
outperforms existing standard THP and RS with linear precoding
schemes.
\end{abstract}
\vspace{0.1cm}
\IEEEpeerreviewmaketitle

\section{Introduction}

Multiple-input multiple-output (MIMO) techniques exploit multipath
propagation by using multiple transmit and receive antennas. MIMO
has become a fundamental part of several communications standards,
such as WiFi and LTE, due to its ability to greatly increase the
capacity and reliability of wireless systems \cite{Lu2014}. A major
research focus over the last decade has been on multi-user MIMO
(MU-MIMO) systems. However, MU-MIMO systems suffer from multi-user
interference (MUI) that can be dealt in the downlink (DL) by
preprocessing the transmit signal at the Base Station (BS) using
precoding algorithms. The quality of the Channel State Information
(CSI) affects the performance of precoding algorithms. However, the
ability to obtain highly accurate CSI at the transmitter (CSIT) is
questionable \cite{JoudehClerckx2016,Clerckx2016}.

Recently, Rate-Splitting (RS), originally developed for the 2-user
SISO interference channel in \cite{Han1981}, has been introduced for
the design of MIMO wireless networks \cite{Clerckx2016}. RS schemes
split the data transmitted from BS to the users into a common
message and private messages. The common message must be decoded by
all users, whereas the private message is decoded only by its
corresponding user. The benefit of RS lies in its capability to
partially decode interference and partially treat interference as
noise, which enables to softly bridge the two extremes of fully
decoding interference and treating interference as noise. As a
consequence, RS provides room for rate and QoS enhancements in a
wide range of network loads (underloaded and overloaded regimes) and
user deployments (with a diversity of channel directions, channel
strengths and qualities of CSIT) over standard schemes such as
MU-MIMO with linear precoding and power-domain Non-Orthogonal
Multiple Access (NOMA) \cite{Clerckx2016,Maoinpress}.

RS has been considered with linear precoding
\cite{JoudehClerckx2016,Maoinpress,Hao2015} using both perfect and
imperfect CSIT. In particular, the sum-rate (SR) maximization problem
using RS and linear precoding in the DL has been investigated. In
\cite{Joudeh2017}, the problem of achieving max-min fairness amongst
multiple co-channel multicast groups has been studied. RS has also
been considered for robust transmissions under bounded CSIT errors
in \cite{Joudeh2016a}. Studies of massive MIMO and MISO networks
using RS strategies have been reported in \cite{Dai2016} and
\cite{Hao2017}, respectively. RS has so far been studied and
optimized using a linear precoding framework. Interestingly, the
potential benefits of RS using nonlinear precoding techniques remain
unexplored in the literature and we aim at filling this gap in this
paper.

The combination of RS and nonlinear precoding is particularly
interesting in the imperfect CSIT setting. Indeed, we know that
nonlinear precoding comes very close to the optimal performance
(sum-rate capacity) of a multi-antenna Broadcast Channel (BC),
achieved by DPC, in the perfect CSIT setting. The sum-rate capacity
of a K-user multi-antenna BC with imperfect CSIT remains an open
problem, even though we know that RS is the key building block to
achieve the optimal Degrees-of-Freedom of a K-user multi-antenna BC
with imperfect CSIT
\cite{JoudehClerckx2016,Piovano2017,Davoodi2016}. From a DoF
perspective, linearly precoded RS is sufficient and any form of
nonlinear precoding combined with RS would not further increase the
DoF. However, from a rate perspective, nonlinear precoding is
beneficial over linear precoding. Hence the combination of nonlinear
precoding and RS is a promising avenue to improve the rate
performance, especially in the imperfect CSIT setting.

In this paper, we consider the design of nonlinearly precoded RS
based on Tomlinson-Harashima Precoding (THP), simply denoted as
THP-RS. In particular, we consider RS with centralized and
decentralized THP structures, where only one user splits its
message, and develop expressions to describe the
signal-to-interference-plus-noise (SINR) ratio and the sum rates
associated with these schemes. Furthermore, we also examine the
benefits of combining RS with Dirty-Paper Coding (DPC). We evaluate
the performance of THP-RS and existing schemes using the sum-rate as
a metric, in perfect and imperfect CSIT.

The rest of the paper is organized as follows. Section II describes
the system model and reviews standard linear precoding, THP and RS
for multiuser MISO systems. Section III details the proposed THP-RS
schemes, whereas the simulations are presented in Section IV.
Finally, Section V concludes the paper.



\section{System Model}

We consider a multiple-input single-output (MISO) BC with $K$ users.
The transmitter delivers a total of $K$ messages to the $K$ users,
with one message intended per user. The BS is equipped with a total
of $N_t$ antennas with $ N_t \geq K \geq 2$, whereas the terminals
of the users are equipped with a single antenna. The transmission
takes place over a channel whose parameters remain fixed during a
data packet. The channel matrix $\mathbf{H} = [{\mathbf h}_1 \ldots
{\mathbf h}_k \ldots {\mathbf h}_K]$ contains in the $k$th column
the channel vector that connects the BS to user $k$. From this
model, we can express the received signal at the $k$th user by
\begin{equation}
r_k=\mathbf{h}_k^H\mathbf{x}+n_k, \label{eq1}
\end{equation}
where $\mathbf{x}\in\mathbb{C}^{N_t}$ represents the transmitted
signal, $n_k\sim \mathcal{CN}(0,\sigma^2_{n,k})$ is the additive
white Gaussian noise, $(\cdot)^H$ is the Hermitian transpose and
$\mathbf{h}_k\in\mathbb{C}^{N_t}$ is the channel vector for user
$k$. In this work, for simplicity we consider equal noise variance
$\sigma_n^2$ for all users. The SNR is defined as
$\text{SNR}\triangleq E_{tr}/ \sigma_n^2$, where $E_{tr}$ denotes
the total transmitted power. We also consider that $\sigma_n^2$
remains fixed and has a non-zero value in order to avoid
indetermination. This means that a modification of the SNR depends
only on the parameter $E_{tr}$. The model satisfies the transmit
power constraint
$\mathbb{E}\left[\lVert\mathbf{x}\rVert^2\right]\leq E_{tr}$. In
what follows we will review a standard MISO BC with linear precoding
and THP, as well as RS using linear precoding.

\subsection{Standard Linear Precoding}

In a standard MISO BC using linear precoding
\cite{Joham2005,Keke2013,wlgmi}, we consider $K$ messages encoded
into $K$ independent data streams, forming the vector
$\mathbf{s}=\left[s_1,s_2,\dots,s_K\right]^{\text{T}}$, where the
superscript $^T$ denotes transpose. Moreover, we assume that
$\mathbb{E}\left[\mathbf{s}\mathbf{s}^{H}\right]=\mathbf{R_s}=\mathbf{I}$,
with $\mathbf{I}$ the identity matrix. The precoding matrix
$\mathbf{P}\in\mathbb{C}^{N_t\times K}$ maps the symbols to the
transmit antennas. The $k$th column of $\mathbf{P}$ contains the
precoder for user $k$, denoted by $\mathbf{p}_k$. It turns out that
the transmit vector is given by
$\mathbf{x}=\mathbf{P}\mathbf{s}=\sum_{k=1}^K \mathbf{p}_k s_k$.
Taking into account the assumptions made so far, the power
constraint is reduced to
$\text{tr}\left(\mathbf{P}\mathbf{R_s}\mathbf{P}^H\right)\leq
E_{tr}$.

\subsection{Linearly Precoded Rate-Splitting}

RS splits a message into a common part and a private part \cite{Clerckx2016,Han1981}. For
simplicity, we consider that only one user splits its message. The
common part is then encoded into one common stream and the private
parts into $K$ private streams. The receivers share a codebook since
the common message has to be decoded by all the users with zero
error probability. In contrast, each private stream is decoded only
by its corresponding user. This means that each receiver must decode
two data streams, namely the common stream (decoded by all but
intended to only one user) and the private stream (decoded by its
respective user). The common stream is first decoded and all private
messages are considered as interference and treated as noise. Then
we use successive interference cancellation (SIC) to subtract the
contribution of the common stream from the received signal,
enhancing the detection of the private stream. At the end, the
message sent via the private stream is decoded. When a user decodes
its private stream, it treats the other private streams as noise.
The strength of RS is its ability to adjust the content and the
power of the common message to control how much interference should
be decoded by all users (through the common message) and how much interference is treated as
noise.

RS can be viewed mathematically as a non-orthogonal unicast and multicast transmission strategy given the superimposed transmission of common and private messages. However, 
conventional multicast messages are intended and decoded by all the users while the common message of RS is decoded by all users but is intended to one (or a subset) of the users. Its presence enables the decoding of part of the MUI and treating the remaining part of the interference as noise.

Splitting one message creates $K+1$ streams, which modifies
the vector of data symbols to
$\mathbf{s}_{\text{RS}}=\left[s_c,s_1,s_2,\dots,s_K\right]^{\text{T}}$,
where $s_c$ is used to designate a symbol of the common stream. A common
precoder $\mathbf{p}_c \in \mathbb{C}^{N_t}$ is added to the first
column of $\mathbf{P}$, from which we obtain
$\mathbf{P}_{\text{RS}}=\left[\mathbf{p}_c,\mathbf{P}\right]$. The
transmitted signal is expressed by
\begin{equation}
\mathbf{x}=\mathbf{p}_c s_c+\sum_{k=1}^K\mathbf{p}_k s_k.
\end{equation}
The total transmit power is allocated partially to the private precoders and the common precoder. For uncorrelated inputs, the transmit power constraint is given by $\lVert \mathbf{p}_c \rVert^2 + \sum_{k=1}^K{\lVert \mathbf{p}_k\rVert^2}\leq E_{tr}.$
Setting $\lVert \mathbf{p}_c \rVert^2$ to zero is equivalent to
allocating no power to the common stream, i.e., the system performs
no RS and the transmit signal is reduced to the standard linear precoding\footnote{Power-domain NOMA is also a particular case of RS whenever the entire message of a given user is encoded into the common message \cite{Maoinpress}.}. 
Power allocation can be carried out to satisfy several system
requirements such as maximizing the SR or achieving a specific
QoS. Given a channel state, the average receive power at
the $k$th terminal can be written as


\begin{equation}
T_{c,k} \triangleq  \mathbb{E}\left[\lvert r_k\rvert^2\right]= \lvert\mathbf{h}_k^H\mathbf{p}_c\rvert^2+I_{c,k},
\end{equation}
with
\vspace{-0.2cm}
\begin{align}
I_{c,k}=&\lvert\mathbf{h}_k^H\mathbf{p}_k\rvert^2+I_{k}, &
I_{k}=&\sum_{\substack{i=1\\i\neq k}}^K\lvert\mathbf{h}_k^H\mathbf{p}_i\rvert^2+\sigma_n^2,
\end{align}
where $I_{c,k}$ and $I_{k}$ correspond the interference-plus-noise
power when decoding the common and the $k$th private message, respectively.

\subsection{Tomlinson-Harashima Precoding}

THP is a nonlinear preprocessing technique employed at the transmit
side. A standard THP algorithm implements three filters, the
feedback filter $\mathbf{B}\in \mathbb{C}^{N_t\times K}$, the
feedforward filter $\mathbf{F}\in\mathbb{C}^{N_t\times N_t}$ and the
scaling matrix $\mathbf{G}\in\mathbb{C}^{N_t\times N_t}$. The
feedback filter deals with the multiuser interference by
successively subtracting the interference from the current symbol.
The matrix $\mathbf{B}$ has a lower triangular structure, whereas
the feedforward filter enforces the spatial causality. The scaling
filter assigns a coefficient or weight to each stream of data, which
means $\mathbf{G}$ is a diagonal matrix.

There are two general THP structures in the literature
\cite{Tomlinson1971,Harashima1972,Windpassinger2004,ZuLamareHaardt2014,Zhang2014},
namely the centralized THP (cTHP) and the decentralized THP (dTHP).
The main difference between these structures is that the scaling
matrix $\mathbf{G}$ is placed at the transmitter for the cTHP,
whereas for dTHP the same matrix is located at the receiver. These
THP algorithms are implemented by performing an LQ decomposition on
the channel matrix, i.e.,
$\mathbf{\mathbf{H}=\mathbf{L}\mathbf{Q}}$. The THP filters are then
defined as follows:
\begin{align}
\mathbf{F}=&\mathbf{Q}^H,\\
\mathbf{G}=&\text{diag}\left(l_{11},l_{22},\dots,l_{KK}\right)^{-1},
\end{align}
\begin{equation}
\mathbf{B}^{\left(d\right)}=\mathbf{G}\mathbf{L} ~~~~~~~{\rm and} ~~~~~
\mathbf{B}^{\left(c\right)}=\mathbf{L}\mathbf{G},
\end{equation}
where $\mathbf{B}_d$ and $\mathbf{B}_c$ correspond to the feedback
filter for dTHP and cTHP, respectively. The received signal vector for each structure is obtained by stacking up the received
signal of each user $r_k$ and is given by
\begin{align}
\mathbf{r}^{\left(d\right)}=&\frac{1}{\beta^{\left(d\right)}} \mathbf{G}\left(\mathbf{H}\beta^{\left(d\right)}\mathbf{F}\mathbf{{w}}+\mathbf{n}\right),\\
\mathbf{r}^{\left(c\right)}=&\frac{1}{\beta^{\left(c\right)}}\left(\mathbf{H}\beta^{\left(c\right)}\mathbf{F}\mathbf{G}\mathbf{{w}+\mathbf{n}}\right),
\end{align}
where $\beta^{\left(d\right)}\approx\sqrt{\frac{E_{tr}}{K}}$ and
$\beta^{\left(c\right)}\approx\sqrt{\frac{E_{tr}}{\sum_{k=1}^K\left(1/l_{k,k}^2\right)}}$
are the scaling factors used to fulfil the transmit power constraint.

The transmitted symbol $w_k$ of each user is successively generated
as
\begin{equation}
w_k=s_k-\sum_{i=1}^{k-1}b_{k,i}w_i.
\end{equation}
However, this process increases the amplitude of $w_k$. A modulo
operation is therefore applied in order to reduce the amplitude of
the symbol to the boundary of the modulation. The modulo processing
is equivalent to adding a perturbation vector $\mathbf{d}$ to the
transmit data $\mathbf{s}$, i.e.,
$\mathbf{v}=\mathbf{s}+\mathbf{d}$. Mathematically, the feedback
processing is equivalent to an inversion operation over the matrix
$\mathbf{B}$. Then, we have
\begin{equation}
\mathbf{{w}}=\mathbf{B}^{-1}\mathbf{v}=\mathbf{B}^{-1}\left(\mathbf{s}+\mathbf{d}\right).
\end{equation}
We can simplify the received signal using the expressions of the
filters, which leads us to
\begin{align}
\mathbf{r}^{\left(d\right)}=&\mathbf{v}+\frac{1}{\beta^{\left(d\right)}}\mathbf{G}\mathbf{n}\\
\mathbf{r}^{\left(c\right)}=&\mathbf{v}+\frac{1}{\beta^{\left(c\right)}}\mathbf{n}
\end{align}
THP introduces a power loss and a modulo loss in the system. The
former comes from the energy difference between the original
constellation and the transmitted symbols after precoding. The
latter is caused by the modulo operation. Both losses can be
neglected for analysis purposes and for moderate and large
modulation sizes so that the power of $\mathbf{v}$ is approximated
by that of $\mathbf{s}$ \cite{ZuLamareHaardt2014}. We remark that
the covariance matrices of the error
$\boldsymbol{\Psi}^{\left(d\right)}=\mathbb{E}\left[\left(\mathbf{r}^{\left(d\right)}-\mathbf{v}\right)\left(\mathbf{r}^{\left(d\right)}-\mathbf{v}\right)^H\right]$
and
$\boldsymbol{\Psi}^{\left(c\right)}=\mathbb{E}\left[\left(\mathbf{r}^{\left(c\right)}-\mathbf{v}\right)\left(\mathbf{r}^{\left(c\right)}-\mathbf{v}\right)^H\right]$
directly affect the performance of the precoders. Since
$\Psi^{\left(c\right)}_{k,k}>\Psi^{\left(d\right)}_{k,k}$
\cite{ZuLamareHaardt2014}, dTHP outperforms cTHP. However, cTHP
requires less complex receivers than dTHP.


\subsection{THP Rate Analysis}
Here, we consider that the power loss is measured by the factor $1/
\lambda>1$, i. e., $\mathbf{R}_{\bf v}= E[{\bf v} {\bf v}^H] =
\lambda^{-1}\mathbf{I}$ \cite{SungMcKay2014}. Now the scaling factors
$\beta^{\left(d\right)}$ and $\beta^{\left(c\right)}$ have to be
multiplied by $\sqrt{\lambda}$. Then, the SINR for the $k$th user is
given by
\begin{align}
\gamma_k^{\left(d\right)}=&\frac{\lambda E_{tr} l^2_{k,k}}{K \sigma^2_n}, &
\gamma_k^{\left(c\right)}=&\frac{
\lambda E_{tr}}{\sigma_n^2\sum\limits_{i=1}^{K}\left(1/l^2_{i,i}\right)}.\label{SR perfect CSIT}
\end{align}
Assuming Gaussian distributed codebooks, the corresponding
instantaneous rates for the $k$th user are given by
\begin{align}
R_k^{\left(d\right)}=&\log_2\left(1+\gamma_k^{\left(d\right)}\right),
&
R_k^{\left(c\right)}=&\log_2\left(1+\gamma_k^{\left(c\right)}\right)\label{THP
rates}.
\end{align}

Let us now consider the imperfect CSIT scenario. Due to the estimation errors, the
channel can be written as
\begin{equation}
\mathbf{H}=\mathbf{\hat{H}}+\mathbf{H}_e,
\end{equation}
where $\mathbf{\hat{H}}$ represents the channel estimate and
$\mathbf{H}_e$ is a random matrix corresponding to the error for
each link. The channel for user $k$ can be written as
$\mathbf{h}_k=\mathbf{\hat{h}}+\mathbf{h}_{e,k}$. Each coefficient
of the error matrix follows a Gaussian distribution, i.e.,
$\sim\mathcal{CN}(0,\sigma^2_{e})$. Because of the errors present in
the CSIT, both algorithms, dTHP and cTHP, can no longer effectively
subtract the interference from other users. Therefore, the received
signal for the conventional THP with imperfect CSIT for both schemes can be expressed as
\begin{align}
\mathbf{r}^{\left(d\right)} =&\mathbf{v}+ \mathbf{G}\mathbf{H}_e\mathbf{F}\mathbf{B}^{-1}\mathbf{v}+\frac{1}{\beta^{\left(d\right)}}\mathbf{G}\mathbf{n},\label{received dthp imperfect CSIT}\\
\mathbf{r}^{\left(c\right)} =&\mathbf{v}
+\mathbf{H}_e\mathbf{F}\mathbf{G}\mathbf{B}^{-1}\mathbf{v}+\frac{1}{\beta^{\left(c\right)}}\mathbf{n}.
\label{received cthp imperfect CSIT}
\end{align}
We can expand equations \eqref{received dthp imperfect CSIT} and
\eqref{received cthp imperfect CSIT} to get the received signal of
each user as described by
\begin{align}
r_k^{d}=&v_k+\frac{1}{l_{k,k}}\mathbf{h}_{e,k}^H\mathbf{p}_k^{\left(d\right)}v_k + \frac{1}{l_{k,k}}\mathbf{h}_{e,k}^H
\sum_{\substack{i=1\\i\neq k}}^K\mathbf{p}_i^{\left(d\right)}v_i+\frac{n_k}{\beta^{\left(d\right)}l_{k,k}},\label{received user dthp imperfect CSIT}\\
r_k^{c}=&v_k+\mathbf{h}_{e,k}^H\mathbf{p}_k^{\left(c\right)}v_k+
\mathbf{h}_{e,k}^H\sum_{\substack{i=1\\i\neq
k}}^K\mathbf{p}_i^{\left(c\right)}v_i+\frac{1}{\beta^{\left(c\right)}} n_k.\label{received
user cthp imperfect CSIT}
\end{align}
Using \eqref{received user dthp imperfect CSIT} and \eqref{received
user cthp imperfect CSIT}, and substituting the value of $\beta$ we
arrive at the following expressions for the SINR of the $k$th user:
\begin{align}
\gamma_k^{\left(d\right)} =&\frac{\lvert 1 +
\frac{1}{l_{k,k}^2}\mathbf{h}_{e,k}^H\mathbf{p}_k^{\left(d\right)}\rvert^2}{\sum\limits_{\substack{i=1\\i\neq
k}}^K\frac{1}{l_{k,k}^2}
\lvert\mathbf{h}_{e,k}^H\mathbf{p}_i^{\left(d\right)}\rvert^2+\frac{K\sigma_n^2}{\lambda E_{tr}l_{k,k}^2}},\label{imperfect CSIT cTHP Rate}\\
\gamma_k^{\left(c\right)}
=&\frac{\lvert 1 +
\mathbf{h}_{e,k}^H\mathbf{p}_k^{\left(c\right)}\rvert^2}{\sum\limits_{\substack{i=1\\i\neq k}}^K
\lvert\mathbf{h}_{e,k}^H\mathbf{p}_i^{\left(c\right)}\rvert^2+\frac{\sigma^2_n\sum\limits_{j=1}^{K}\left(1/l^2_{j,j}\right)}{\lambda E_{tr}}}. \label{imperfect CSIT dTHP Rate}
\end{align}
Finally, the respective instantaneous rates are found using
\eqref{THP rates}. In order to assess the performance of the
proposed schemes, we adopt the Ergodic Sum-Rate (ESR) $R_{ESR}$ as a
performance metric to average out the effects of CSIT errors and guarantee the rates are achievable \cite{JoudehClerckx2016}. To calculate the
ESR, we first calculate the Average SR (ASR) $\bar{R}_k$,
which is the average performance computed for a given channel
estimate with respect to the errors \cite{JoudehClerckx2016}. The Sample Average
Approximation (SAA) method \cite{2009} is used to approximate the
solution of the stochastic ASR problem. Consider $M$ i.i.d
realizations of the error matrix $\mathbf{H}_e$ where
$\mathbf{H}_e^{(m)}$ denotes the $m$th realization. Then,
$R_k^{\left(m\right)}$  is the rate associated to the $m$th
realization and the $k$th user. Using a SAA approach we have that
the ASR is given by
\begin{equation}
\bar{R}_k=\mathbb{E}\left[ R_k\vert \mathbf{\hat{H}}\right]\approx\frac{1}{M}\sum_{m=1}^M R^{\left( m\right)},\label{ASR}
\end{equation}
The ergodic rate is taken from the expected
value of the ASR over multiple channel estimates, leading to
\begin{equation}
R_{ESR}=\mathbb{E}\left[\sum_{k=1}^K\bar{R}_k\right]\label{ESR}
\end{equation}

\section{Tomlinson-Harashima Precoded Rate-Splitting}

In this section, we present the proposed THP-RS schemes and develop
expressions to compute the SINR and the sum-rate of these schemes.
The main motivation of THP-RS is to improve the sum-rate performance
beyond that achieved by RS with linear precoding and to exploit RS to make THP schemes more robust against imperfect CSIT. The latter is especially
important because imperfect CSIT tends to affect more adversely THP
than linear precoding due to the interference cancellation
\cite{Windpassinger2004,ZuLamareHaardt2014}. Note that due to the
power loss and the modulo loss, THP techniques do not achieve the
performance of DPC \cite{SungMcKay2014}. However, THP is significantly less complex than
DPC.

\subsection{Proposed THP-RS with perfect CSIT}

Let us now investigate whether RS can be combined with THP to
further reduce the gap with DPC in the perfect CSIT setting. We
split the message of one user, linearly precode the common stream
and use THP to precode the private streams. Mathematically, the
transmitted signal is given by
$\mathbf{x}=\left[\mathbf{p}_c,\mathbf{P}^{\left(\text{THP}\right)}
\right]\left[s_c,\mathbf{v}^{\text{T}}\right]^{\text{T}}$. Taking
into account that cTHP and dTHP define the structure of
$\mathbf{P}^{\left(\text{THP}\right)}$, we get
\begin{align}
\mathbf{P}^{\left(\text{d}\right)}=&\beta^{\left(d\right)}\mathbf{F}\mathbf{B}^{\left(d\right)^{-1}},\label{dTHP Precoder}\\
\mathbf{P}^{\left(\text{c}\right)}=&\beta^{\left(c\right)}\mathbf{F}\mathbf{G}\mathbf{B}^{\left(c\right)^{-1}}.\label{cTHP Precoder}
\end{align}
The transmitted signal is then rewritten as
\begin{equation}
\mathbf{x}=\mathbf{p}_c s_c+\sum_{i=1}^K \mathbf{p}_i v_i,\label{RS transmit signal}
\end{equation}
where $\mathbf{p}_i$ is the $i$th column of the precoder defined in
\eqref{dTHP Precoder} or \eqref{cTHP Precoder}, depending on the
structure adopted. Note that both scaling factors $\beta$ are
modified since part of the power is assigned to the common precoder,
leading to
$\beta^{\left(d\right)}\approx\sqrt{\frac{\lambda\left(E_{tr}-\lVert\mathbf{p}_c\rVert^2\right)}{K}}$,
and
$\beta^{\left(c\right)}\approx\sqrt{\frac{\lambda\left(E_{tr}-\lVert\mathbf{p}_c\rVert^2\right)}{\sum_{i=1}^{K}\left(1/l^2_{i,i}\right)}}$.
Then, the received signals of the proposed THP-RS schemes are
described by
\begin{align}
\mathbf{r}^{\left(\text{rs-d}\right)}=&\frac{1}{\beta^{\left(d\right)}}\mathbf{G}\mathbf{H}\mathbf{p}_c s_c+\mathbf{v}+\frac{1}{\beta^{\left(d\right)}}\mathbf{G}\mathbf{n},\\
\mathbf{r}^{\left(\text{rs-c}\right)}=&\frac{1}{\beta^{\left(c\right)}}\mathbf{H}\mathbf{p}_c s_c+\mathbf{v}+\frac{1}{\beta^{\left(c\right)}}\mathbf{n}.
\end{align}
At the $k$th user we have
\begin{align}
r_k^{\left(\text{rs-d}\right)}=&\frac{1}{\beta^{\left(d\right)}l_{k,k}}\mathbf{h}_k^H\mathbf{p}_c s_c+v_k+\frac{n_k}{\beta^{\left(d\right)}l_{k,k}},\\
r_k^{\left(\text{rs-c}\right)}=&\frac{1}{\beta^{\left(c\right)}}\mathbf{h}_k^H\mathbf{p}_c s_c+v_k+\frac{1}{\beta^{\left(c\right)}}{n_k}.
\end{align}
From the last equation we obtain the SINR for the common message of
the $k$th user:
\begin{align}
\gamma_{c,k}^{\left(d\right)}=&\frac{K\lvert\mathbf{h}_k^H\mathbf{p}_c\rvert^2}{\lambda l^2_{k,k}\left(E_{tr}-\lVert\mathbf{p}_c\rVert^2\right)+ K\sigma_n^2},\\
\gamma_{c,k}^{\left(c\right)}=&\frac{\sum\limits_{i=1}^{K}\left(1/l^2_{i,i}\right)\lvert\mathbf{h}_k^H\mathbf{p}_c\rvert^2}{\lambda E_{tr}+\sigma_n^2\sum\limits_{i=1}^{K}\left(1/l^2_{i,i}\right)}.
\end{align}
The instantaneous rates for user $k$ can be obtained by
$R_{c,k}^{\left(d\right)}=\log_2\left(1+\gamma_{c,k}^{\left(d\right)}\right)$
and
$R_{c,k}^{\left(c\right)}=\log_2\left(1+\gamma_{c,k}^{\left(c\right)}\right)$
for dTHP and cTHP, respectively. The common rate is set to $R_c =
\min R_{c,k}$ to ensure that all users can decode the message. After
decoding the common message, the receiver subtracts it from the
received signal. The SINR expressions for the private messages are
\begin{align}
\gamma_{p,k}^{\left(d\right)}=&\frac{\lambda l^2_{k,k}\left(E_{tr}-\lVert\mathbf{p}_c\rVert^2 \right)}{K \sigma^2_n}, \\
\gamma_{p,k}^{\left(c\right)}=&\frac{
\lambda \left(E_{tr}-\lVert\mathbf{p}_c\rVert^2\right)}{\sigma_n^2\sum\limits_{i=1}^{K}\left(1/l^2_{i,i}\right)},
\end{align}
which are similar to \eqref{SR perfect CSIT}. However, the value of
$\gamma_k$ is reduced due to the power assigned to the common
message. It follows that the instantaneous rate for the private
message is given by \eqref{THP rates}. At the end, the sum-rates for
the RS system can be expressed as
\begin{align}
R^{\left(d\right)}=&R_{c}^{\left(d\right)} +\sum_{k=1}^K
R_k^{\left(d\right)}, &
R^{\left(c\right)}=&R_{c}^{\left(c\right)}+\sum_{k=1}^K
R_k^{\left(c\right)}.\label{RS inst Rate}
\end{align}

\subsection{Proposed Rate-Splitting THP with imperfect CSIT}

In this section, we consider THP-RS under imperfect CSIT. Using
\eqref{RS transmit signal} we can express the received signal as
follows:
\begin{align}
\mathbf{r}^{\left(\text{rs-d}\right)}=&\frac{1}{\beta^{\left(d\right)}}\mathbf{G}\mathbf{H}\mathbf{p}_c s_c+\mathbf{v}+\mathbf{G}\mathbf{H}_e\mathbf{F}\mathbf{B}^{-1}\mathbf{v}+\frac{1}{\beta^{\left(d\right)}}\mathbf{G}\mathbf{n},\\
\mathbf{r}^{\left(\text{rs-c}\right)}=&\frac{1}{\beta^{\left(c\right)}}\mathbf{H}\mathbf{p}_c s_c+\mathbf{v}+\mathbf{H}_e\mathbf{F}\mathbf{G}\mathbf{B}^{-1}\mathbf{v}+\frac{1}{\beta^{\left(c\right)}}\mathbf{n}.
\end{align}
From the last equation we can obtain the received signal at each
user equipment, which is given by
\begin{align}
r_k^{\left(\text{rs-d}\right)}=&\frac{\beta^{\left(d\right)^{-1}}}{l_{k,k}}\mathbf{h}_k^H\mathbf{p}_c s_c+v_k+\frac{1}{l_{k,k}}\mathbf{h}_{e,k}^H\sum_{\substack{i=1}}^K\mathbf{p}_i^{\left(d\right)}v_i+\frac{\beta^{\left(d\right)^{-1}}n_k}{l_{k,k}},\\
r_k^{\left(\text{rs-c}\right)}=&\frac{1}{\beta^{\left(c\right)}}\mathbf{h}_k^H\mathbf{p}_c s_c+v_k+\mathbf{h}_{e,k}^H\sum_{\substack{i=1}}^K\mathbf{p}_i^{\left(c\right)}v_i+\frac{ n_k}{\beta^{\left(c\right)}}.
\end{align}
Then, the SINR for the common message can be computed by the
following:
\begin{align}
\gamma_{c,k}^{\left(d\right)}=&\frac{\lvert\mathbf{h}_k^H\mathbf{p}_c\rvert^2/\beta^{\left(d\right)}}{\lvert
l_{k,k}+
\mathbf{h}_{e,k}^H\mathbf{p}_k^{\left(d\right)}\rvert^2+\sum\limits_{\substack{i=1\\i\neq
k}}^K
\lvert\mathbf{h}_{e,k}^H\mathbf{p}_i^{\left(d\right)}\rvert^2+\sigma_n^2/\beta^{\left(d\right)}},\\
\gamma_{c,k}^{\left(c\right)}=&\frac{\lvert\mathbf{h}_k^H\mathbf{p}_c\rvert^2/\beta^{\left(c\right)}}{\lvert
1+\mathbf{h}_{e,k}^H\mathbf{p}^{\left(c\right)}_k\rvert^2+\sum\limits_{\substack{i=1\\i\neq
k}}^K
\lvert\mathbf{h}_{e,k}^H\mathbf{p}_i^{\left(c\right)}\rvert^2+\sigma_n^2/\beta^{\left(c\right)}}.
\end{align}
The rate of the private messages can be calculated with equations \eqref{imperfect CSIT dTHP Rate} and \eqref{imperfect CSIT cTHP Rate}. The transmit power should be changed to $E_{tr}-\lVert\mathbf{p}_c\rVert^2$, as explained before for the perfect CSIT case. The resulting sum rate is computed with \eqref{RS inst Rate}. Then
the ASR and ESR can be found with \eqref{ASR} and \eqref{ESR},
respectively. The rates for ZF-DPC-RS approximation based on
\cite{CaireShamai2003} with imperfect CSIT can be
obtained by using the previous expressions and by neglecting the
power loss and the modulo loss.

\section{Simulations}

In this section we evaluate the performance of the proposed THP-RS
schemes using zero-forcing (ZF) filters and compare them with
existing techniques. We consider a MISO BC channel with $4$ transmit
antennas and $4$ users, where each user is equipped with a single
antenna. The inputs follow a Gaussian distribution with variance
$\sigma_s^2=1$. We also consider additive white Gaussian noise and
flat fading Rayleigh channels scenario, where all the users experience the same SNR.
 The ASR was calculated using a total
of $100$ error matrices for each estimated channel. Then the ESR was
obtained averaging over $50$ independent channel estimates. The
precoder for the common message was obtained using a singular value
decomposition (SVD) of the channel matrix
($\mathbf{H}=\mathbf{USV}$), i. e., $\mathbf{p}_c=\mathbf{V}(:,1)$.
A percentage of the power from the private precoders has been
assigned to the common precoder. The power assigned to the common
precoder was found through an optimization procedure, while the
remaining power was uniformly allocated across the private
precoders.

\begin{figure}[htb]
\begin{centering}
\includegraphics[scale=0.37]{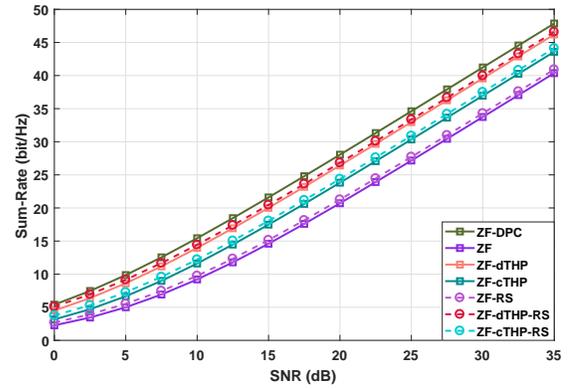}
\par\end{centering} \vspace{-0.75em}
\caption{Sum rate performance with RS and perfect
CSIT.}\label{PerfectCSIT}
\end{figure}

\begin{figure}[htb]
\begin{centering}
\includegraphics[scale=0.37]{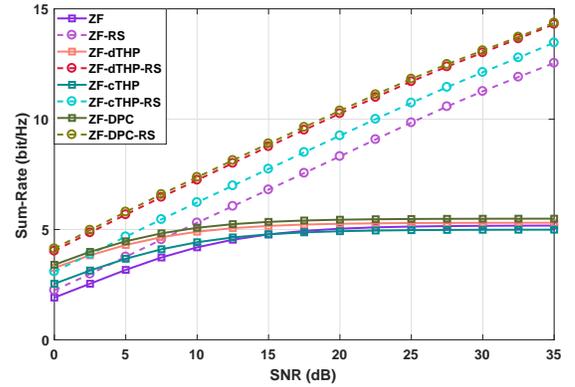}
\par\end{centering} \vspace{-0.75em}
\caption{Sum rate performance with RS and imperfect
CSIT.}\label{ImperfectCSIT}
\end{figure}

Figs. \ref{PerfectCSIT} and \ref{ImperfectCSIT} illustrate the
results for the precoding algorithms with perfect and imperfect
CSIT, respectively. We consider a power loss factor of
$\lambda=0.75$ for the THP structures. For imperfect CSIT we used a
fixed error variance equal to $0.2$. The results show that the
proposed THP-RS schemes outperform previously reported THP and
linear schemes. RS-based schemes only offer a small gain over
standard schemes with perfect CSIT, as illustrated in Fig.
\ref{PerfectCSIT}, whereas those gains are substantial in the
presence of imperfect CSIT, which corroborates the sum-rate results
in the literature for linearly precoded RS schemes\footnote{Recall
however that simulation results hold only for sum-rates, under
uniform power allocation among private streams and with users
experiencing similar channel strengths. If we allow optimization of
the power across private streams, or choose a weighted sum-rate or
experience user channel strength disparity, RS-based schemes can
provide larger gains, as detailed in \cite{Maoinpress}.}. The
results also show that the ZF-DPC approximation obtains the highest
sum rates, as expected, followed by THP and linear schemes.
Specifically among RS-based schemes, the ZF-DPC-RS
\cite{CaireShamai2003} obtains the best result followed by dTHP-RS,
cTHP-RS and linearly precoded RS. Note that for the ZF-DPC
implemented here, we considered uniform power allocation among the
streams. The performance advantage of dTHP structures over cTHP ones
is explained by the error covariance matrix previously presented,
and by the use of more complex receive filters at the users, whereas
cTHP only employs filters at the transmitter which translates into
lower complexity \cite{ZuLamareHaardt2014}. The curves obtained for
imperfect CSIT exhibit saturation due to the fact that the variance
of the CSIT errors does not scale with the SNR
\cite{JoudehClerckx2016,Davoodi2016}. This is expected for THP
schemes due to error propagation associated with imperfect CSIT.

\begin{figure}[htb]
\begin{centering}
\includegraphics[scale=0.37]{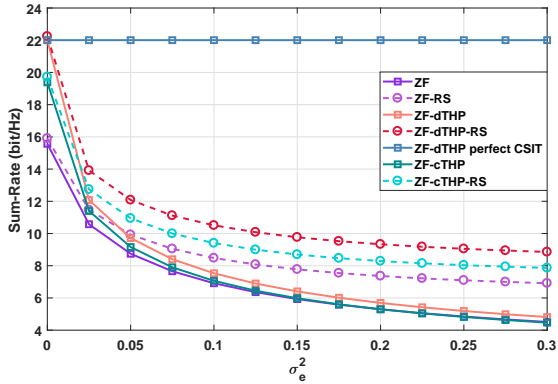}
\par\end{centering} \vspace{-0.75em}
\caption{Sum rate performance versus channel error
variance}\label{SRvsVarErr}
\end{figure}

In the next example, we consider the sum-rate performance for
different error variances using $SNR = 15$ dB, as illustrated in
Fig. \ref{SRvsVarErr}. The results show that RS increases robustness
for all error variances. In particular, the proposed dTHP-RS scheme
achieves the highest sum rate, followed by cTHP-RS and linearly
precoded RS. The sum rates achieved by dTHP-RS can be up to $25\%$
higher than cTHP-RS and RS, whereas they can be up to $100 \%$
higher than those achieved by non RS-based schemes. This highlights
the robustness of RS schemes against imperfect CSIT for a wide range
of scenarios.

\begin{figure}[htb]
\begin{centering}
\includegraphics[scale=0.37]{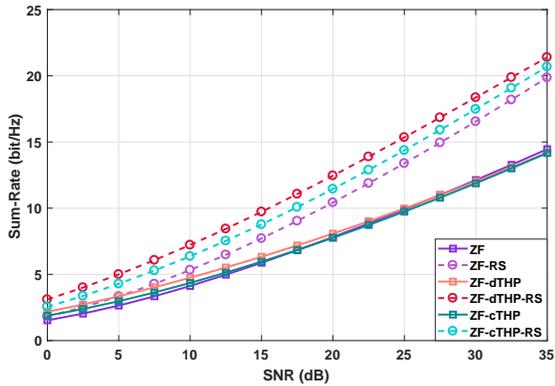}
\par\end{centering} \vspace{-0.75em}
\caption{Sum rate performance with RS, imperfect CSIT and
$\alpha=0.6$.}\label{CSITquality}
\end{figure}

In the last example, we consider that the variance of the error
scales with the SNR $\left(\sigma_e^2=E_{tr}^{-\alpha}\right)$. The
curves obtained in Fig. \ref{CSITquality} have been computed with
$\alpha=0.6$. The results indicate that THP-RS schemes are more
robust than standard THP schemes and achieve higher sum rates than
linear schemes. It can be noticed that the slope achieved by
RS-based schemes is significantly higher than that associated with
non RS-type approaches, which corroborates the robustness shown in
Fig. \ref{SRvsVarErr} and the superiority of RS in terms of DoF
\cite{JoudehClerckx2016,Piovano2017}.

\section{Conclusion}

In this paper we have proposed THP-RS schemes and derived SINR and
sum-rate expressions to evaluate their performance with perfect and
imperfect CSIT. Moreover, we have also examined the sum-rate
performance of ZF-DPC with and without RS for perfect and imperfect
CSIT. Simulation results have shown that the proposed THP-RS schemes
can achieve higher sum rates than those of existing THP and linear
schemes, and are more robust against imperfect CSIT than standard
THP schemes.






%
\bibliographystyle{IEEEtran}
\bibliography{PrecodingAlgorihtms}
%

%
%

\end{document}